\documentclass[letterpaper, 10 pt, conference]{ieeeconf}  
\usepackage{booktabs}
\usepackage{graphics}
\usepackage{amsmath}
\usepackage{xcolor}
\usepackage{dsfont}
\usepackage{graphicx}
\usepackage[export]{adjustbox}

\usepackage{stfloats}

\DeclareMathOperator*{\minimize}{minimize}
\renewcommand \baselinestretch{0.98}
\usepackage{fullpage}
\usepackage{amssymb,amsmath}\newcommand{\RR}{\mathbb R}
\usepackage{morefloats}

\usepackage{subcaption}
\usepackage[labelformat=parens,labelsep=quad,skip=3pt]{caption}
\usepackage{graphicx}

\IEEEoverridecommandlockouts                              

\title{\LARGE \bf
A Graph-Constrained Changepoint Learning Approach for Automatic QRS-Complex Detection
}

\author{Atiyeh Fotoohinasab,  Toby Hocking, and Fatemeh Afghah,
\thanks{A. Fotoohinasab, T. Hocking and F. Afghah are with the School of Informatics, Computing and Cyber Systems at Northern Arizona University.}
}

 \renewcommand{\baselinestretch}{.92}
\begin{document}

\maketitle
\thispagestyle{empty}
\pagestyle{empty}

\begin{abstract}
This study presents a new viewpoint on ECG signal analysis by applying a graph-based changepoint detection model to locate R-peak positions. This model is based on a new graph learning algorithm to learn the constraint graph given the labeled ECG data. The proposed learning algorithm starts with a simple initial graph and iteratively edits the graph so that the final graph has the maximum accuracy in R-peak detection. We evaluate the performance of the algorithm on the MIT-BIH Arrhythmia Database. The evaluation results demonstrate that the proposed method can obtain comparable results to other state-of-the-art approaches. The proposed method achieves the overall sensitivity of \textit{Sen} = 99.64\%, positive predictivity of \textit{PPR} = 99.71\%, and detection error rate of \textit{DER} = 0.19.
\end{abstract}


\section{INTRODUCTION}
The electrocardiogram (ECG) is a quasi-periodic biomedical signal, which serves as the most commonly used non-invasive tool in the diagnosis of cardiovascular diseases. One cardiac cycle in a typical ECG signal is identified by arrangements of P, QRS-complex, and T waveforms as well as PQ and ST segments. In most automated ECG analysis tools, correct R-wave detection is of great importance as it is used for detecting other ECG fiducial points and is served a significant criterion for the diagnosis of many heart arrhythmias. 
Nevertheless, efficient R-peak detection is still a challenging task due to the time-variant waveform morphology caused by noise corruption or a specific cardiac condition.

There are various methods in the literature that have been proposed to detect R-peak in ECG signal \cite{beraza2017comparative}. However, these methods mainly suffer from some critical drawbacks that limit their implementations in practical applications. First, the non-stationary morphology of the QRS complex can lead to the misdetection of R-peak, especially in some determinant morphological patterns concerning certain life-threatening heart arrhythmias. Considering this limitation, it is apparent that incorrect identification of other waves can occur subsequent to the incorrect detection of R-peaks. Second, the performance of most R-peak detection algorithms highly depends on a preprocessing step to remove the impact of different types of noise contaminating the ECG signal. However, in real-time data processing and ambulatory care settings, preprocessing-based algorithms are less effective. Several studies have utilized deep learning techniques to detect the ECG waveforms motivated by the high performance of deep learning methods in various classification tasks \cite{mousavi2020single}. However, the problem with deep learning-based algorithms is that they need large scale datasets to train the algorithm and often suffer from the imbalanced class problem \cite{8683140}.

In \cite{fotoohinasab2020graph}, we introduced a new class of graph-based statistical models based on optimal changepoint detection models, named graph-constrained changepoint detection (GCCD) model to locate R-peak in the ECG signal. 
This work was the first study performed to apply a changepoint detection model for extracting ECG fiducial points. 
The GCCD model is constrained to a graph under which prior biological knowledge of the signal is taken into account in order to accomplish the segmentation task. However, the performance of the previous GCCD model depends on the choice of the constraint graph, which is manually defined by an expert with prior knowledge. To tackle this issue, in this paper, we propose a new graph-based changepoint detection model that learns the structure of the constraint graph in the labeled ECG data. 
It is worth mentioning that, like the previous model, the proposed method does not require any preprocessing step as it leverages the sparsity of changepoints to denoising the signal as well as detecting abrupt changes.  

The rest of the paper is organized as follows. In the next section, we describe the proposed Graph-Constrained Changepoint Detection model and its application in the localization of R-peak. Section \ref{sec:Experimental Studies} provides a description of the dataset used in this study and a discussion of the results as well as a comparison between the performance of the proposed algorithm and other state-of-the-art algorithms. Finally, \ref{sec:CONCLUSIONS and DISCUSSION} summarizes this research work and its contributions.

\section{Methodology}
 \label{sec:GCCD}
In \cite{fotoohinasab2020graph}, we developed an initial GCCD model based on a manually defined constraint graph using regarding the labeled ECG data. This GCCD model extracted the R-peaks in an ECG signal by representing the periodic non-stationary ECG signal as a piecewise-stationary time series with constant mean values per segment \cite{fotoohinasab2020graph}. To automate the ECG segmentation task, we propose a new model, which takes a raw ECG signal and an initial graph structure as its inputs and yields the onset/offset and the mean of desired segments regarding the structure of the learned constraint graph. 
The model architecture is composed of two steps, including training and detection. The training phase is utilized for optimizing the structure of the constraint graph. It takes the raw ECG signal and an initial graph structure as the inputs and tries to find an optimum graph structure that can minimize the label errors over the training set. Then, the changepoint detection model extracts the R-peaks in the raw ECG record subject to the trained graph from the previous step. In the following sections, we describe the details of various parts of the proposed model.  

\subsection{Graph-Constrained Changepoint Detection Model}

ECG waves can be considered as abrupt up or down changes over time per cardiac cycle. Thus, we employ the optimal changepoint detection model introduced in \cite{hocking2017log} to localize R-peak positions in the ECG signal. 
In this framework, the prior biological knowledge about the expected sequence of changes are specified in a graph as the model constraints. Then, a functional pruning dynamic programming algorithm is used to compute the globally optimal model (mean, changes, hidden states) in fast log-linear $O(N\log N)$ time.  

The constraint graph can be defined with a directed graph $G=(V,E)$, where a vertex set $V\in\{1,\dots,|V|\}$ represents the hidden states/segments (not necessarily a waveform), and the edge set $E\in\{1,\dots,|E|\}$ represents the expected changes between the states/segments. Each edge $e\in E$ encodes the following associated data based on the prior knowledge about the expected sequences of changes:
 \begin{itemize}
    \item The source $\underline v_e\in V$ and target $\overline v_e\in V$ vertices/states for a changepoint $e$ from $\underline v_e$ to $\overline v_e$.
    \item A non-negative penalty constant $\lambda_e\in\RR_+$ which is the cost of changepoint $e$.
    \item A constraint function $g_e:\RR\times\RR\rightarrow\RR$ which defines the possible mean values before and after each changepoint $e$. If $m_i$ is the mean before the changepoint and $m_{i+1}$ is the mean after the changepoint, then the constraint is $g_e(m_i,m_{i+1})\leq 0$. These functions can be used to constrain the direction (up or down) and/or the magnitude of the change (greater/less than a certain amount).
\end{itemize}

Given the input signal $Y=\{y_1,\dots,y_n\}$ and the directed graph $G=(V,E)$, the problem of finding changepoints ~$\mathbf c$, segment means $\mathbf m$, and hidden states $\mathbf s$ can be mathematically stated as follows; 

\begin{align}
  \label{eq:op-c}
  & \minimize_{
  \substack{
  \mathbf m\in\RR^N,\, \mathbf s\in V^N\\
   \mathbf c\in\{0,1,\dots,|E|\}^{N-1}\\
  }
  } \ \ 
      \sum_{i=1}^N \ell(m_i, z_i) + \sum_{i=1}^{N-1} \lambda_{c_i} \\
        \text{s. t\ \ } &\ \text{no change: }c_i = 0 \Rightarrow m_i=m_{i+1} ~\& ~s_i=s_{i+1} \, \label{eq:nochange-constraint}\\ \nonumber
    &\ \text{change: }c_i \neq 0 \Rightarrow g_{c_i}(m_i,m_{i+1})\leq 0 ~~\& \\
        &(s_i,s_{i+1})=(\underline v_{c_i},\overline v_{c_i}).\label{eq:change-constraint}
\end{align}
where the segment means $\mathbf m$, the hidden states $\mathbf s$, and the changepoints~$\mathbf c$ are the optimization variables. Also, the changepoints $c_i$ can be any of the pre-defined edges ($c_i\in\{1,\dots,|E|\}$), or $c_i=0$ which indicates no change (and has no cost, $\lambda_0=0$). The above objective function ~(\ref{eq:op-c}) consists of a data-fitting term $\ell$ and a model complexity term $\lambda_{c_i}$. $\ell$ represents the negative log-likelihood of each data point, and $\lambda_{c_i}$ is a non-negative penalty on each changepoint. 
The constraint function $g_e$ also encodes the expected up/down change and the least amplitude gap between the mean of two states. The constraint~(\ref{eq:nochange-constraint}) enforces the model to keep its current state $s_i=s_{i+1}$ with no change in mean $m_i=m_{i+1}$ when there is no change $c_i=0$. On the other hand, the constraint~(\ref{eq:change-constraint}) forces a change in the mean implied by the constraint function $g_{c_i}(m_i,m_{i+1})\leq 0$, and a change in the state $(s_i,s_{i+1})=(\underline v_{c_i},\overline v_{c_i})$ when there is a change $c_i\neq 0$.

\subsection{Constraint Graph Learning}
\label{sec:Graph Learning}
The constraint graph $G=(V,E)$ in the optimization problem of ~(\ref{eq:op-c}) can be designed manually or in a learning framework. In \cite{fotoohinasab2020graph}, we constructed the constraint graph, including the graph topology and edge information manually using some pre-specified categories for each waveform \cite{tafreshi2014automated}. However, the manual definition of the constraint graph can be inefficient for ECG signal analysis considering the various morphological patterns for each waveform. Besides, the performance of the model depends on the expert knowledge while encoding the prior knowledge into the constraint graph. Therefore, in this work, we propose a graph learning algorithm for learning the constraint graph using the R-peak labels provided by a gold standard.
   
\begin{figure*}[htb]
      \centering
      \includegraphics[width=\linewidth,height=\textheight,keepaspectratio]{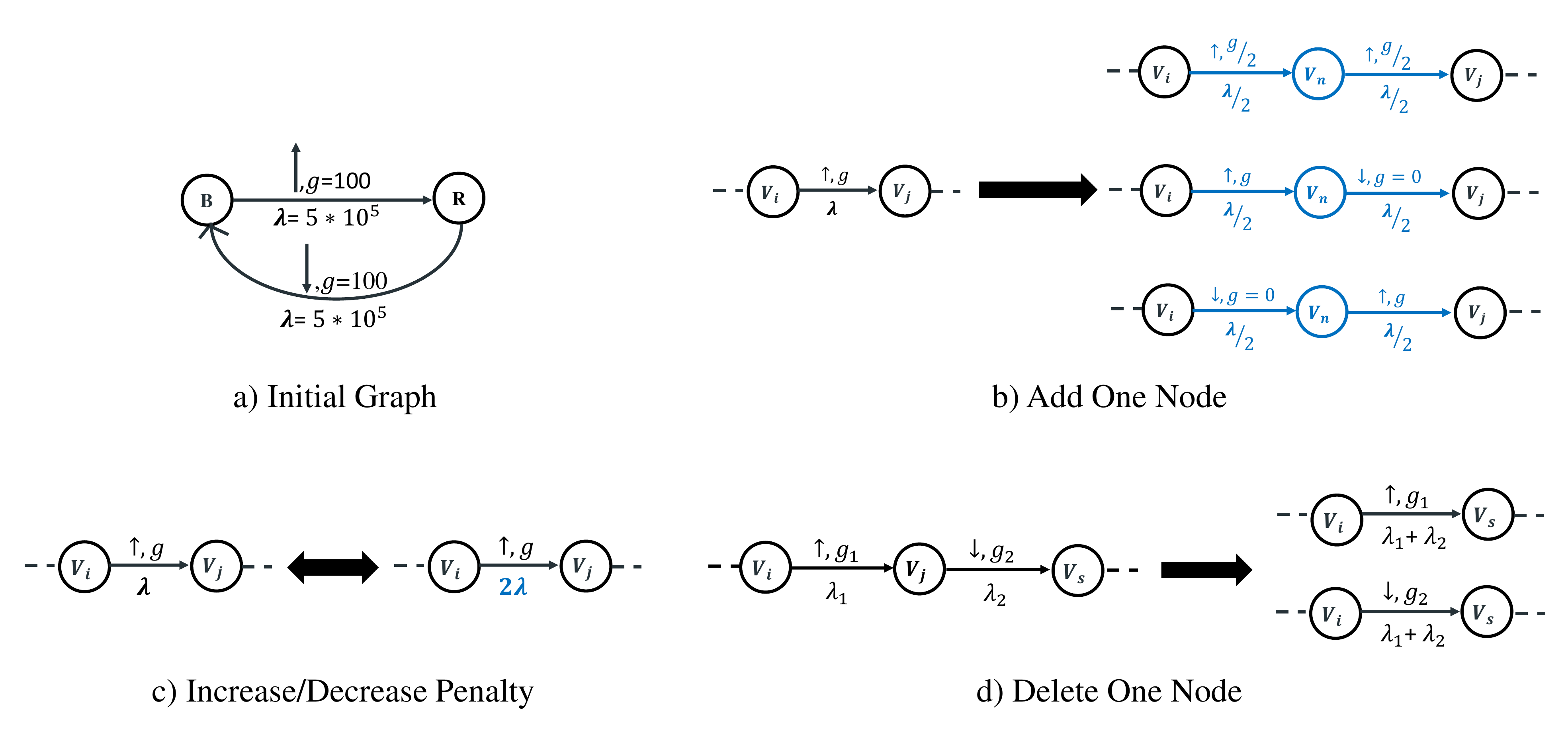}
      \caption{\textbf{a:} The initial constraint graph with two nodes labeled as $B$ and $R$, representing the baseline and the R-peak segments, respectively, in a cycle.
      \textbf{b-d:} Some of the applied graph editing candidates related to the edge $(V_i, V_j)$ with an up change.
      }
      \vspace{-15pt}
      \label{addnode}
   \end{figure*}

The goal of constraint graph learning is to automatically discover the desired graph topology property (i.e., identifying the desired $V$ and $E$ sets in graph $G$) and the information of edges from data. In this study, the edge information consists of the expected up/down change in the segment means, the least amplitude gap between the mean of two states, and a non-negative penalty corresponded to the edge transition. In order to learn the constraint graph, the proposed graph learning algorithm starts with an initial graph, including an initial graph topology and an initial edge information set; and then, iteratively optimizes the graph parameters to find a graph that maximizes the accuracy regarding the given labels. Figure \ref{addnode}a shows the simple initial graph used for the optimization process. It should be noted that the initial edge information is chosen based on the overall results obtained from the manual definition of the constraint graph.   

The algorithm considers 10 editing candidates per edge to heuristically determine the graph candidate set in each iteration. 
Graph editing candidates set in the proposed algorithm consists of three types of adding a node, two types of deleting a node, one type of adding two nodes, and increasing or decreasing the penalty and gap corresponding to an edge. We believe all morphological patterns of the ECG waveforms can be constructed using these editing candidates. Figure \ref{addnode} exemplifies some of the applied graph editing candidates related to the edge $(V_i, V_j)$ with an up change.

\section{Experimental Studies}
\label{sec:Experimental Studies}
\subsection{Dataset}
We applied the publicly available MIT-BIH arrhythmia (MIT-BIH-AR) database to evaluate the proposed model \cite{moody2001impact, goldberger2000physiobank}. The MIT-BIH-AR database contains 48 ECG recordings taken from 47 subjects, and each recording is sampled at 360 Hz for 30 min with 200 samples resolution over a 10 mV range. Each recording consists of two ECG leads, including leads V1, V2, V4, V5, and the modified lead II (MLII). Only the MLII and V5 are used to evaluate the performance of the algorithm. The database is annotated with both RR intervals and heartbeat class information by two or more expert cardiologists independently. We employ the provided annotations for R-wave positions in order to train the model and evaluate its performance. The training and testing sets are also generated by randomly dividing the ECG cycles per records with an approximate ratio of 3:1. 
\subsection{Experimental Results}
\label{sec:Results}
Figure \ref{addnode2} demonstrates the performance of the proposed model in the localization of R-peak for subject 107 from the MIT-BIH-AR dataset. This figure illustrates how the proposed algorithm iteratively edits the simple initial graph to find a graph with maximum accuracy in detecting R-peaks regarding the labels in the training set. The red part of the graph in each iteration shows the modified part of the graph regarding the constraint graph in the previous iteration. It is worth noting that the performance of the proposed algorithm depends on the initial graph structure. Therefore, the models for 5 records out of all records are trained based on a more complex initial graph than the initial graph in \ref{addnode}a. We evaluate the performance of the proposed algorithm using the sensitivity (Sen), positive predictivity rate (PPR), and detection error rate (DER), which are defined as below:

 \begin{figure*}[htb]
      \centering
      \includegraphics[width=\linewidth,height=0.87\textheight,keepaspectratio]{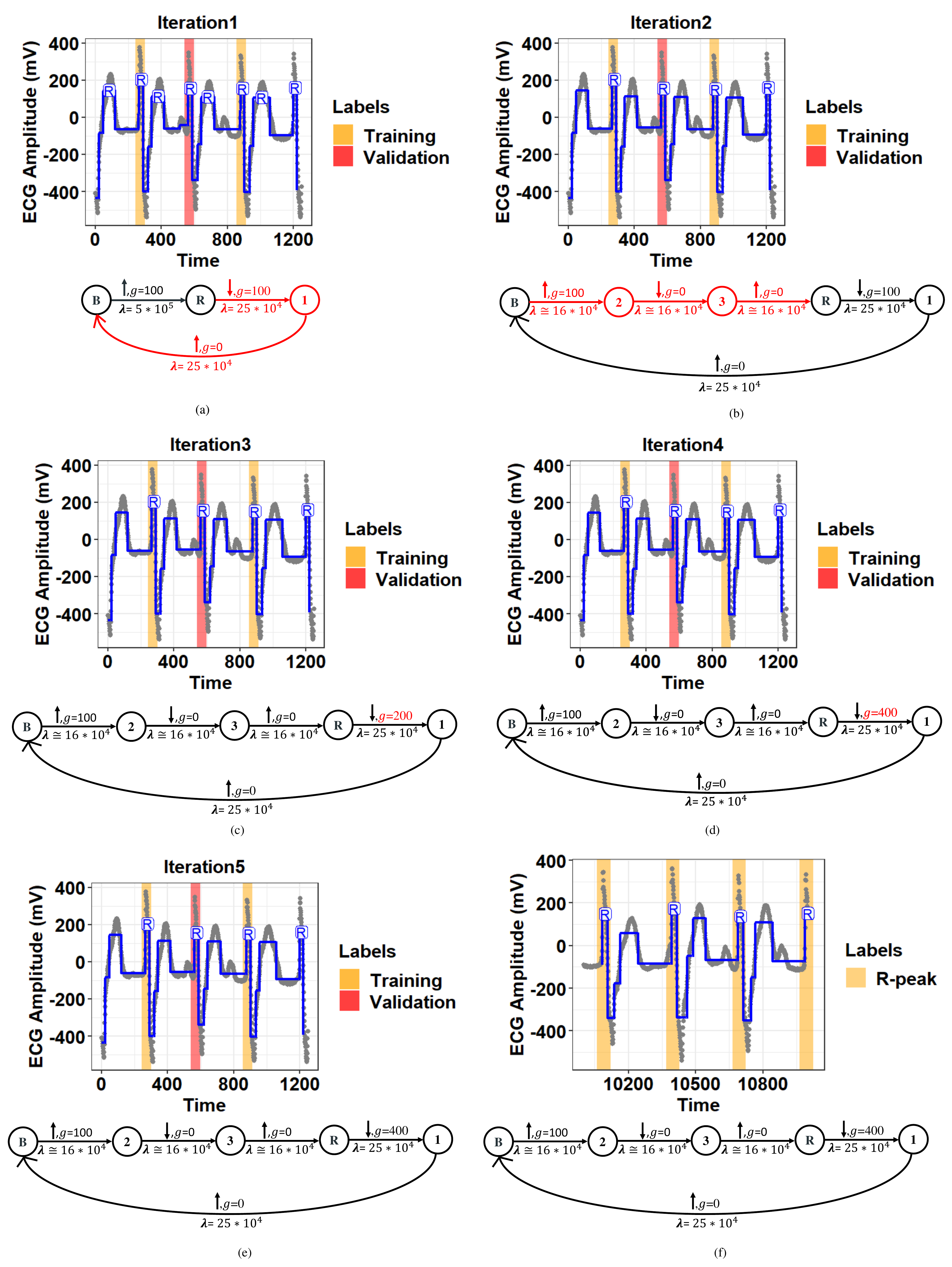}
      \caption{The demonstration of constraint graph optimization using the proposed graph learning algorithm for the subject 107 from the MIT-BIH-AR dataset.
     \textbf{a-e top:} Extracted R-peak positions given the learned constraint graph in each learning iteration. The red and orange coverage bands show the used labels in the training procedure, including both training and validation sets. 
     \textbf{a-e bottom:} The learned constraint graph in each learning iteration. Below each edge $e$, we show the penalty $\lambda_e$, and above, we show the type of change (i.e., up/down) and the gap, which is the minimum magnitude of change. The red part of the graph in each iteration shows the modified part of the graph regarding the constraint graph in the previous iteration. 
      \textbf{f:} Testing the final learned constraint graph in a new window of data. The orange coverage band shows the labels provided by the MIT-BIH-AR dataset.
}
      \vspace{-15pt}
      \label{addnode2}
   \end{figure*}
   
    \begin{figure}[htb]
      \centering
      \includegraphics[height=1.0\textheight,width=1.0\linewidth,keepaspectratio]{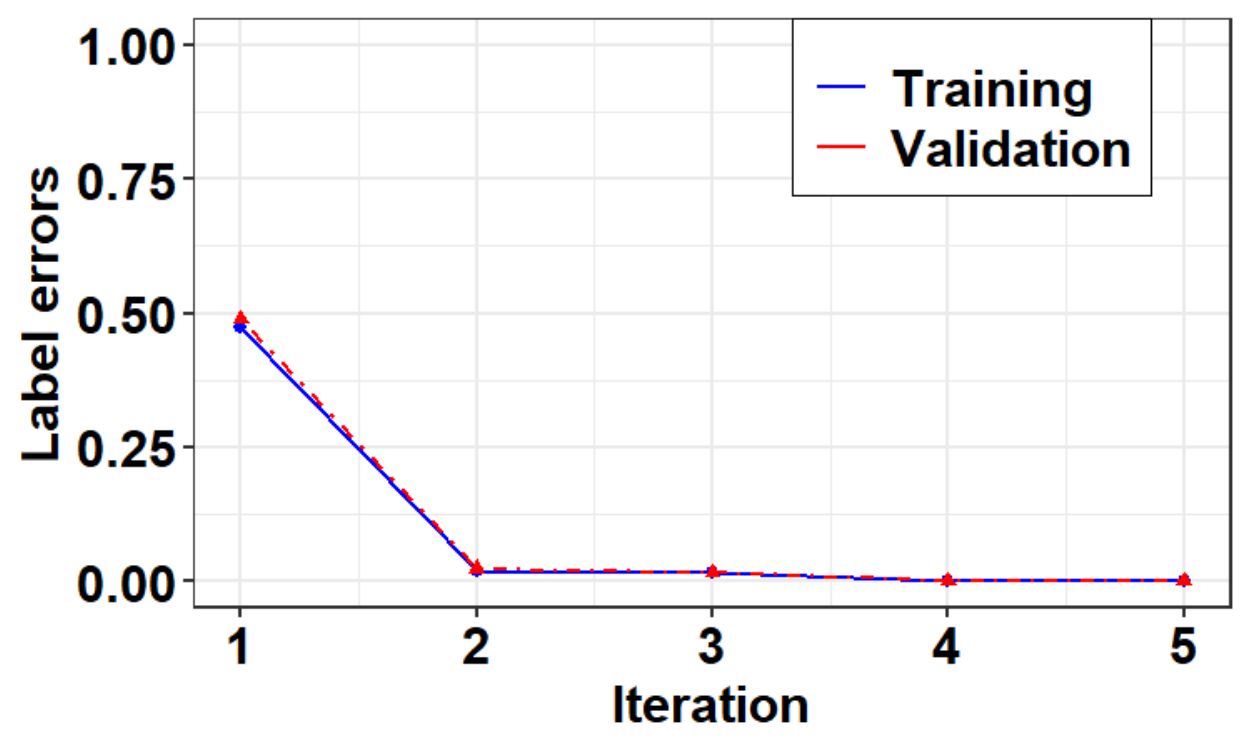}
      \caption{Label errors in detection of R-peak positions regarding the learned constraint graph per 5 learning iterations for subject 107 in the MIT-BIH-AR dataset.}
      \vspace{-15pt}
      \label{TrainingProgress}
   \end{figure}

\begin{align}
\label{eq:sen}
\textit{Sen} (\%) =  \frac{TP}{TP+FN}\times{100}\\
\label{eq:PPR}
\textit{PPR}  (\%) =  \frac{TP}{TP+FP}\times{100}\\
\label{eq:DER}
\textit{DER} (\%) =  \frac{FN+FP}{TP+FN}\times{100}
\end{align}
where TP is true positives, FP is false positives, FN is false negatives, and TN is true negatives. We used $k$-fold cross-validation approach to train and test the proposed model with a $k$ size of 5. Indeed, we divided the dataset into k= 5 folds. Then, for each fold of the 5 folds, one fold is used for evaluating the model, and the remaining 4 folds are used to train the model. In the end, the final result was averaged over all 5 folds. Figure \ref{TrainingProgress} illustrates the training progress over 5 iterations for record 107 from the MIT-BIH-AR dataset, where the Y axis shows the sum of false negative and false positive error rates. 

Table \ref{tab:Comparison_of_methods} represents the R-peak detection success for the proposed algorithm against other state-of-the-art methods. As shown in the table, the proposed algorithm achieves remarkable results, \textit{Sen} = \%99.64, \textit{PPR}  = \%99.71, and \textit{DER} = 0.19, in R-peak detection. We note that here our remarkable results were obtained regardless of a preprocessing step, as opposed to other methods in the literature. We noticed that the records with \textit{Sen} and \textit{PPR}  values lower than 99\% contain multiple different morphological patterns. In these records, more than one optimum graph path is required in the constraint graph in order to detect R-peaks. The performance of the proposed model for such cases can be improved by learning a multi-path constraint graph, which is considered as future work. Besides, learning a multi-path constraint graph can be applied for detecting all ECG waveforms as considering morphological patterns of each waveform leads to having cycles with completely various morphological patterns. 

\begin{table}[ht]
\caption{Comparison of performance of several R-peak detection methods using the MIT-BIH-AR database}
 \centering{
\label{tab:Comparison_of_methods}
\resizebox{1.\linewidth}{!}{ 
\begin{tabular}{cccc}
\toprule
\textbf{Method} & \textbf{\textit{Sen} (\%)} & \textbf{\textit{PPR}  (\%)} & \textbf{\textit{DER} (\%)}\\ 
\midrule
\texttt Park et al. \cite{park2017r} & 99.93 & 99.91 & 0.163\\
\texttt Farashi \cite{farashi2016multiresolution} & 99.75 & 99.85 & 0.40 \\
\texttt Sharma and Sunkaria \cite{sharma2017qrs} & 99.50 & 99.56 & 0.93\\
\texttt Castells-Rufas and Carrabina \cite{castells2015simple} & 99.43 & 99.67 & 0.88\\
\textbf {Proposed Model} & \textbf {99.64} &  \textbf {99.71} & \textbf {0.19}\\
\hline
\end{tabular}}}
\end{table}

\section{Conclusion}
\label{sec:CONCLUSIONS and DISCUSSION}
In this paper, the task of R-peak detection in the ECG signal is carried out based on a new viewpoint using a changepoint detection model. The proposed algorithm uses a changepoint detection algorithm constrained to a graph in which prior biological knowledge of the expected changepoints per cardiac cycle is encoded. We propose an algorithm for learning the constraint graph in labeled data. The proposed learning algorithm starts from an initial graph and optimizes the structure of the constraint graph as well as the edges information to find a model that can maximize the detection accuracy. The results provided in this paper demonstrates that changepoint detection models constrained to a graph are promising approaches for detecting ECG waveforms. The proposed learning algorithm can be advanced to learn a multi-path constraint graph in order to detect all ECG waveforms and find the overall ECG morphological patterns per cardiac cycle.

\addtolength{\textheight}{-12cm}   


\bibliographystyle{plain}
\bibliography{references}
\end{document}